\documentclass{aa}
\usepackage{graphicx}
\usepackage{natbib}
\usepackage{txfonts}

\begin{document}

\title{Eight new 2+2 doubly eclipsing quadruple systems detected}

\author{Zasche,~P.~\inst{1},
        Henzl,~Z.~\inst{2,3},
        Merc, J.~\inst{1},
        K\'ara, J.~\inst{1},
        Ku\v{c}\'akov\'a,~H.~\inst{1,3,4,5}
        }

\offprints{Petr Zasche, \email{zasche@sirrah.troja.mff.cuni.cz}}

 \institute{
  $^{1}$ Charles University, Faculty of Mathematics and Physics, Astronomical Institute, V~Hole\v{s}ovi\v{c}k\'ach 2, CZ-180~00, Praha 8, Czech Republic\\
  $^{2}$ Hv\v{e}zd\'arna Jaroslava Trnky ve Slan\'em, Nosa\v{c}ick\'a 1713, Slan\'y 1, 274 01, Czech Republic \\
  $^{3}$ Variable Star and Exoplanet Section, Czech Astronomical Society, Fri\v{c}ova 298, 251 65 Ond\v{r}ejov, Czech Republic \\
  $^{4}$ Astronomical Institute, Academy of Sciences, Fri\v{c}ova 298, CZ-251 65, Ond\v{r}ejov, Czech Republic\\
  $^{5}$ Research Centre for Theoretical Physics and Astrophysics, Institute of Physics, Silesian University in Opava, Bezru\v{c}ovo n\'am. 13, CZ-746 01, Opava, Czech Republic\\
 }

\titlerunning{Eight doubly eclipsing quadruples detected}
\authorrunning{Zasche et al.}

 \date{Received \today; accepted ???}

\abstract{We studied eight new doubly eclipsing stellar systems. We found that they are all rare
examples of quadruple systems of 2+2 architecture, where both inner pairs are eclipsing binaries. Until
now, such a configuration had only been proven for dozens of systems on the whole sky. We enlarged this
rare group of systems with four stars in the Small Magellanic Cloud (SMC) galaxy and four brighter
stars on the northern sky. These analysed systems are the following: OGLE SMC-ECL-2339 (both eclipsing
periods of 0.72884 days and 3.39576 days; mutual orbital period of 5.95 years); OGLE SMC-ECL-3075
(1.35890 d, 2.41587 d, 9.75 yr); OGLE SMC-ECL-4756 (0.91773 d, 2.06047 d, 4.34 yr); OGLE SMC-ECL-6093
(0.90193 d, 2.03033 d, 31.2 yr); GSC 01949-01700 (0.24058 d, 0.75834 d, 21.7 yr); ZTF
J171602.61+273606.5 (0.36001 d, 4.51545 d, 19.5 yr); WISE J210935.8+390501 (0.33228 d, 3.51575 d, 1.9
yr); and V597 And (0.46770 d, 0.35250, 20.4 yr). These systems constitute a rare selection of W UMa
stars among the doubly eclipsing quadruples. For all of the systems, new dedicated observations were
obtained as well. V597 And is definitely the most interesting system for several reasons: (1) the
system is the brightest in our sample; (2) it is a rare quintuple (2+2)+1 system; and (3) it is also
closest to the Sun. It yielded the predicted angular separation of the two components of 57 mas, which
is probably within the detection limits for modern, high-angular-resolution techniques.}

\keywords {stars: binaries: eclipsing -- stars: fundamental parameters} \maketitle

\section{Introduction} \label{intro}

\medskip
During the last few years, studies have shown that the 'doubly eclipsing nature' (i.e. two periodic
eclipsing signals coming from one point source on the sky) is probably much more common in stellar
populations than originally thought. The first such example, named V994 Her, was found by
\cite{2008MNRAS.389.1630L}; however, since then quite an extensive collection of similar candidates
have been detected. This is true namely due to the large photometric surveys, automated telescopes and
satellites (see e.g. \citealt{2024MNRAS.527.3995K} or \citealt{2022A&A...664A..96Z}). Nowadays, we know
more than 900 such doubly eclipsing candidate systems.

However, the reason why we should still consider these systems as only candidates is the fact that the
pure detection of {two sets of eclipses} coming from one point source on the sky cannot be taken as
proof that these really constitute the bound quadruple system. For solid proof, one needs some
additional information. Owing to their typical lower brightness, there exists no spectroscopy for most
of them; moreover, the pairs were not resolved into the double. Therefore, an easy and straightforward
method would be the detection of eclipse-timing variations (hereafter ETVs) for both pairs on their
mutual orbit. If we are able to collect an adequately large compilation of photometry and derive times
of eclipses for both pairs, these would behave in opposite manners, and we would definitely be dealing
with a real 2+2 quadruple. We used this approach in our previous studies
(\citealt{2023A&A...675A.113Z}, and \citealt{2022A&A...659A...8Z}), proving the quadruple nature of
several such systems. However, it should be said that the number of such confirmed 2+2 quadruples is
still very limited, and only a small fraction of the candidates were analysed in detail. Hence, any new
contribution to the topic would be welcome.

A great benefit of such systems of four stars in one gravitationally bound system is that they put
rather strict limitations on the light-curve (hereafter LC) solution and modelling of such a system.
Both inner pairs must share the same age, same initial chemical composition, same distance, and so on.
All of these are strict conditions that should be taken into account when constructing a proper model
of both eclipsing pairs.

Moreover, the need for deeper analyses of such triple and quadruple systems with one, two, or even
three eclipsing binaries \citep{2021AJ....161..162P} shows that their origin or formation mechanism has
not yet been explained satisfactorily. There still exist competing theories involving the close
encounters, disc fragmentation, and small-N-body dynamics (see e.g. \citealt{2001IAUS..200...33W},
\citealt{2021Univ....7..352T}, and \citealt{2021ApJ...917...93K}). Finally, such complex multiple
systems are ideal astrophysical laboratories for studying the dynamical effects like the nodal
precession, eccentricity variations, apsidal motion due to the Kozai-Lidov cycles, or the three-body
dynamics in general  (\citealt{2013MNRAS.435..943P}, \citealt{2022MNRAS.515.3773B}).

\medskip

\begin{table*}[t]%[b]
  \caption{Basic information about the systems.}  \label{systemsInfo}
%  \centering
  \scalebox{0.85}{
  \begin{tabular}{c c c c c c c}\\[-3mm]
\hline \hline\\[-3mm]
  Target name                     &  Other name            &      TESS      &     RA      &     DE       & Mag$_{max}$ $^{\star}$ &  Temperature/sp.type       \\
                                  &                        & identification &  [J2000.0]  & [J2000.0]    &            &       information $^{\star\star}$                                      \\
 \hline
 \object{OGLE SMC-ECL-2339}       & OGLE SMC-SC6 67760      & TIC 614595747 & 00 52 05.35 & -72 46 05.79 &  17.91 (V) &  T$_{eff} = 15123.2$ K  \\  %Vmag dle OGLE: Pawlak2016: 2016AcA....66..421P
 \object{OGLE SMC-ECL-3075}       & OGLE SMC-SC7 4316       & TIC 613102402 & 00 54 46.84 & -73 17 24.42 &  17.87 (V) &  T$_{eff} = 11558.4$ K  \\  %Vmag dle OGLE: Pawlak2016: 2016AcA....66..421P
 \object{OGLE SMC-ECL-4756}       & OGLE SMC-SC10 28914     & TIC 182731575 & 01 03 38.78 & -72 13 51.33 &  15.41 (V) &  T$_{eff} = 20776.6$ K  \\  %Vmag dle OGLE: Pawlak2016: 2016AcA....66..421P
 \object{OGLE SMC-ECL-6093}       & OGLE SMC121.3 1863      & TIC 631046788 & 01 24 24.89 & -73 13 25.36 &  16.73 (V) & sp.type B1.5V$^\sharp$  \\  %sp typ   %Vmag dle OGLE: Pawlak2016: 2016AcA....66..421P
 \object{GSC 01949-01700}         & CRTS J085514.9+293656   & TIC 126276576 & 08 55 14.88 & +29 36 57.03 &  14.13 (V) &  T$_{eff} = 4883.6$ K   \\  %Vmag dle UCAC4
 \object{ZTF J171602.61+273606.5} & UCAC4 589-057416        & TIC 257647120 & 17 16 02.62 & +27 36 06.59 &  13.59 (V) &  T$_{eff} = 5960.1$ K   \\  %Vmag dle UCAC4
 \object{WISE J210935.8+390501}   & 2MASS J21093589+3905016 & TIC 166026283 & 21 09 35.89 & +39 05 01.70 &  13.09 (V) &  T$_{eff} = 5499.1$ K   \\  %Vmag dle UCAC4
 \object{V597 And}                & 2MASS J23060406+4835248 & TIC 252646185 & 23 06 04.06 & +48 35 24.90 &  11.86 (V) &  $B_p-R_p = 1.0159$ mag \\  %Vmag dle GSC 2.4.2 > 2008AJ....136..735L
 \hline
\end{tabular}}\\
 {\small Notes: $^\star$ Out-of-eclipse magnitude, $V_{mag}$ taken from OGLE \citep{2016AcA....66..421P}, UCAC4 \citep{2013AJ....145...44Z}, or Guide Star Catalog \citep{2008AJ....136..735L}.  $^{\star\star}$ Effective temperature taken from the Gaia DR3 catalogue \citep{2023A&A...674A...1G}. $^\sharp$ \citep{2019A&A...625A.104R}. }
\end{table*}

\section{Selected stars}

Our system-selection method is quite straightforward and is part of our long-term effort. We are still
collecting the data for several (dozens of) interesting systems both in the northern and southern
hemispheres. When an adequate number of data points are collected, we proceed to the more detailed
analysis of the particular star system. All of the targets were our new discoveries, and their basic
information is summarised in Table \ref{systemsInfo}. There, one can see the name of the particular
star, its position on the sky, and some magnitude and temperature estimates. The latter should be only
taken as a rough estimate with large uncertainty.

What should be noted here is the fact that most of the systems presented contain at lest one binary of
a contact-like (or near contact) LC shape. This is still quite rare, as one can easily see from
existing papers, since most of the detected, doubly eclipsing quadruples (via ETVs, or spectroscopy)
still belong to the group of detached systems. This is quite a natural consequence of the topic; that
is, the detached systems are generally easier to analyse, their LCs are not too blended for LC
disentangling, and their spectral lines are typically not overly blended with each other.

What is clearly seen in Table \ref{systemsInfo} is that we are dealing with two different types of
stars. Four stars from the Small Magellanic Cloud (hereafter SMC) galaxy are part of the early-type
stars with high temperatures, while the four others are the northern-hemisphere stars detected in the
TESS data and are of a later spectral type that is cooler and smaller than our Sun.

\section{Data used for the analysis}

Owing to the relatively low brightness of the stars, we only used the photometric data for our whole
analysis. The best quality photometry is provided by the TESS satellite \citep{2015JATIS...1a4003R}. It
provides us with the uninterrupted data sequence of 27 days in one TESS sector. For different stars,
several sectors of data are sometimes available covering a few years.

Apart from the TESS data, the older photometric archives also were used for extracting the reliable
data for each of the stars. This was not easy in cases where the amplitude of the photometric variation
is small. The other older photometry is almost unavailable (for the southern-sky stars), especially for
the fainter targets (see Table \ref{systemsInfo}), while for the northern-sky stars pair B sometimes
suffers from larger scatter in the ETV diagrams (due to the poor quality LC for pair B). A
colour-coding scheme is also given below in our Fig. \ref{FigLCOC} for distinguishing between the
different data sources. These were on the northern sky mainly the following surveys:
 ASAS-SN \citep{2014ApJ...788...48S,2017PASP..129j4502K},
 Atlas \citep{2018AJ....156..241H},
 WASP \citep{2006PASP..118.1407P},
 ZTF \citep{2019PASP..131a8003M},
 NSVS \citep{2004AJ....127.2436W},
 CRTS \citep{2017MNRAS.469.3688D}, and
 KWS \citep{KWS}.

For the stars in the SMC galaxy, we used the OGLE data obtained in filter $I$
\citep{2013AcA....63..323P}. These data were used in its III and IV phases, sometimes accompanied from
phase II when available. The MACHO photometry \citep{1997ApJ...486..697A} was also used when available.
Only for one system in the SMC (OGLE SMC-ECL-4756) was it possible to also extract useful TESS data for
the analysis and use these data for both A and B pairs. For the rest of the SMC targets, their
magnitudes are too low for TESS to provide reliable photometry.

Aside from these freely available data, we also used our own data obtained especially for this study.
The new data obtained in recent years at different observatories are the following (plotted in red in
Figure \ref{FigLCOC} below):
\begin{itemize}
    \item The 1.54 m Danish telescope on La Silla observatory in Chile, equipped with a CCD camera, data obtained usually in $R$, and $I$ filters.
    \item The 65 cm telescope at Ond\v{r}ejov observatory, Czech Republic, using a G2-MII CCD camera equipped with an $R$ photometric filter.
    \item The 30 cm telescope at a private observatory in Velt\v{e}\v{z}e u Loun, Czech Republic, equipped with an MII G2-8300 CCD camera.
\end{itemize}

\medskip

\section{Analysis}

For the LC model, we used the best available dataset, which was the TESS one (the best sector with the
shortest cadence data) for the northern targets, and the OGLE ($I$) one for the southern SMC targets.
We carried out the LC analysis on these data using the {\sc PHOEBE} program
\citep{2005ApJ...628..426P}.

The whole analysis followed this procedure. A preliminary LC fit of the more dominant pair was done,
producing the residuals for the analysis of the other pair. The derived LC template was then used for
our AFP method \citep{2014A&A...572A..71Z} to derive the times of eclipses. With these eclipse times,
better ephemerides of the binary were derived. With such an ephemerides, the LC was modelled again.
With a better LC fit, the residuals were re-computed. Such an approach was used several times
iteratively until the individual fitting steps provided reasonable stable results.

We usually started with the assumption of equal masses (i.e. mass ratio 1.0) and the equal luminosities
of both pairs (i.e. third light  fraction 50\%). Then, the third light was freed from constraints, as
was also the mass ratio for some of the stars with better data and larger out-of-eclipse variations.

\medskip

\section{Results}

In this section, we focus on the individual systems presented in our analysis. However, stars and their
analyses are only briefly described since for most of them our study still represents their first
publication and the method used was nearly the same for all of them.

 \begin{figure*}
 \centering
  \includegraphics[width=1.015\textwidth]{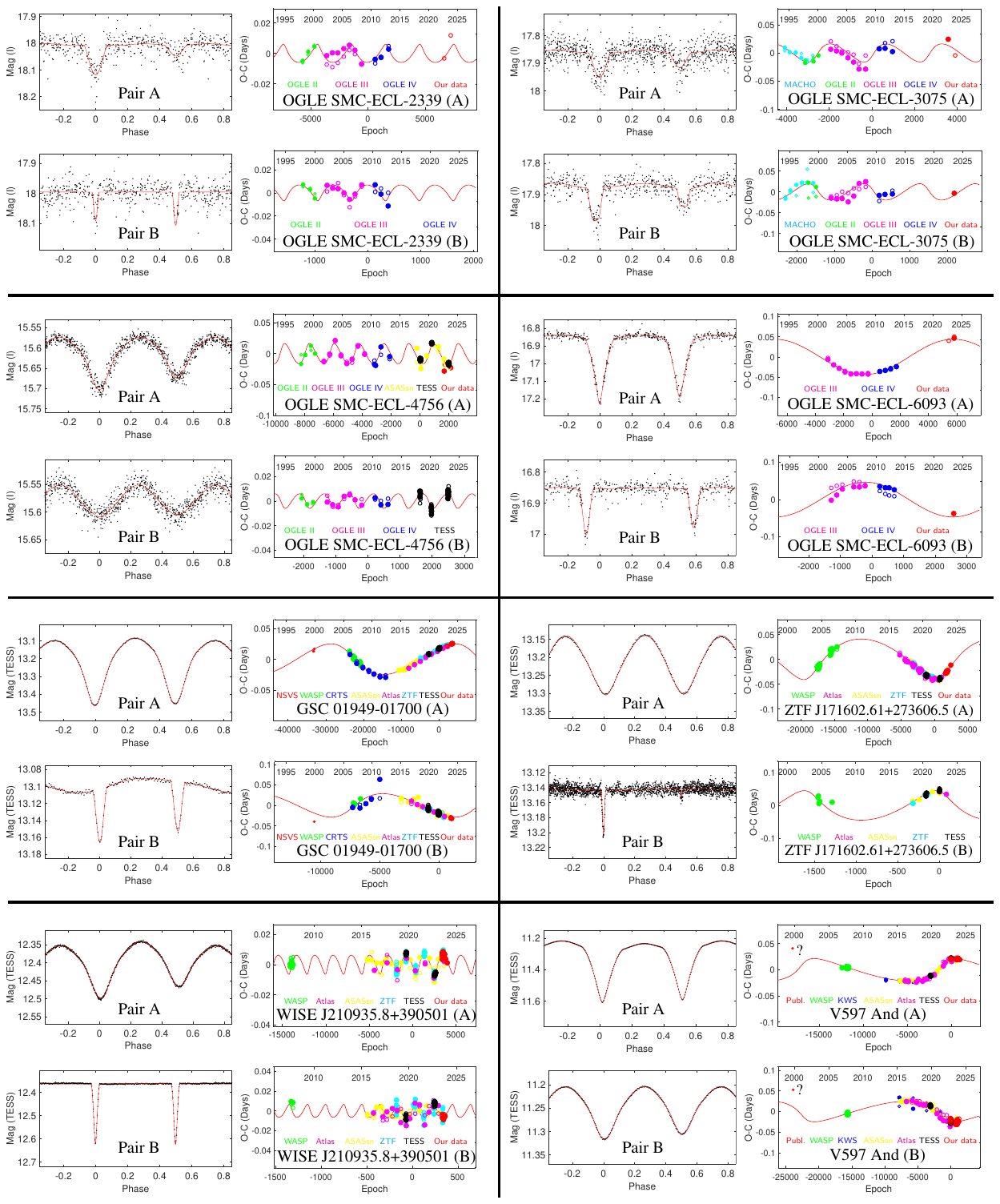}% }
  \caption{Fits for the LC and ETV data for both pairs for all of our analysed systems. { Full dots denote the primary eclipses, open circles the secondary ones. Different sources of data points are distinguished by different colours.}}
  \label{FigLCOC}
 \end{figure*}

\subsection{OGLE SMC-ECL-2339}

The first star studied was OGLE SMC-ECL-2339, which is part of the Small Magellanic Cloud galaxy. This
star had never been studied in detail, probably due to its low brightness. It is the faintest star in
our sample. The star was not recognised as a doubly eclipsing system before, and its only period is
listed as 0.7288 days \citep{2016AcA....66..421P}.

Due to our limited knowledge about the star, many assumptions have to be made. However, our analysis of
both the LCs of pairs A (0.728836 d) and B (3.39576 d), together with the complete ETV analysis of both
pairs, can be seen in Figure \ref{FigLCOC}. We plot\ the disentangled LCs from OGLE data (I filter) for
both pairs A and B and the period variations of both pairs. It is obvious that these ETVs behave in
opposite manners, confirming a quadruple 2+2 nature. Both inner pairs are circular. Parameters of our
fits are given in Tables \ref{TabLC} and \ref{TabETV}.

\subsection{OGLE SMC-ECL-3075}

The next system under our analysis was OGLE SMC-ECL-3075, which is somewhat similar to the previous one
considering its faintness and depth of eclipses. The only detected period by \cite{2016AcA....66..421P}
is 1.35887 days.

Our detailed study revealed that the system also shows the second eclipsing period of about 2.41586
days for pair B. This pair is also slightly eccentric (eccentricity about 0.105; long term apsidal
motion about 113 years). Besides the OGLE III and IV data, the older OGLE II and MACHO
\citep{2007AJ....134.1963F} data are also available. Thanks to these older data, it was much easier to
detect the long-term period variation of both A and B pairs, as plotted in Figure \ref{FigLCOC}.

\subsection{OGLE SMC-ECL-4756}

The system named OGLE SMC-ECL-4756 is the brightest among the SMC stars in our sample. However, it had
not been studied before, and its second eclipsing period is presented here for the first time.

Our analysis (see Fig.\ref{FigLCOC}) shows that the second eclipsing pair B has a lower amplitude and longer period than pair A (for these reasons, our attempts to observe pair B were not successful at all). Both pairs show a contact-like shape of the LC, but the pair B is only marginally eclipsing. There was a surprisingly high difference between the amplitudes from the ETV analyses of A and B, indicating rather different masses of both pairs. However, our LC analysis led to rather comparable third-light fractions. The reason for this discrepancy remains a mystery. This is the only SMC system for which we also used the TESS data for the ETV analysis.  % pair B is near contact, but still slightly eclipsing.  both A+B circular, and ogle2 as well

\subsection{OGLE SMC-ECL-6093}

The last SMC system was OGLE SMC-ECL-6093, which is the only SMC system observed spectroscopically. Its
spectral type was published as B1.5V \citep{2019A&A...625A.104R}. No other detailed information about
the star was found in the published literature.

The analysis in Figure \ref{FigLCOC} led to the parameters given in Tables \ref{TabLC} and
\ref{TabETV}. Here, the amplitude of the ETV for both pairs is in good agreement with the detected
third-light values for A and B. Pair B was found to be eccentric (eccentricity 0.284; apsidal period of
about 70 years). This system shows the longest mutual A-B period in our sample; it is over 30 years
long, but it is still affected by large uncertainties.

\subsection{GSC 01949-01700}

The first presented system from our Galaxy is GSC 01949-01700. This is probably the system with the
latest spectral type classification according to the GAIA DR3 information \citep{2023A&A...674A...1G}.

We found both pairs of A and B LCs to be asymmetric. Therefore, we introduced a hypothesis of spots for both pairs for a proper modelling of the LC shape. Both pairs show circular orbits and are short-period pairs. There was quite a large difference between the resulting third-light fractions, while the amplitudes from the ETV analysis led to more similar pairs.  % EL3 dohromady necelych 100% luminosity
Here, the predicted angular separation of the double to be resolved (according to GAIA DR3 parallax)
was computed to be of about 47 mas, which is probably feasible with the current high-angular-resolution
technique. However,  its low brightness would be problematic.

\begin{table*}%[h!]
\caption{Derived parameters for the two inner binaries A and B.}
 \label{TabLC}
% \scriptsize
 \tiny
  \centering \scalebox{0.73}{
\begin{tabular}{c | c | c | c | c | c | c | c | c  }
   \hline\hline %\noalign{\smallskip}
 \multicolumn{1}{c|}{System } &  OGLE SMC-ECL-2339    & OGLE SMC-ECL-3075 & OGLE SMC-ECL-4756 & OGLE SMC-ECL-6093 &  GSC 01949-01700 & ZTF J171602.61+273606.5 & WISE J210935.8+390501  & V597 And  \\ \hline
  \multicolumn{1}{c|}{ }      &  \multicolumn{7}{c}{{\sc p a i r \,\, A}} \\ \hline
  $i$ [deg]                   &  76.96 $\pm$ 1.75     & 70.81 $\pm$ 1.06  &  65.28 $\pm$ 0.73 &  88.35 $\pm$ 0.70 & 72.76 $\pm$ 0.28 &  60.71 $\pm$ 0.62   &  63.83 $\pm$  0.25  & 84.29 $\pm$ 0.32  \\
 $q=\frac{M_2}{M_1}$          &  1.00 (fixed)         &  1.00 (fixed)     &  1.00 (fixed)     &  1.00 (fixed)     &  0.79 $\pm$ 0.09 &   1.08 $\pm$ 0.05   &   0.97 $\pm$ 0.02   &  1.00 (fixed)     \\
 $T_1$ [K]                    &  15123 (fixed)        & 11558 (fixed)     &  20777 (fixed)    &  24500 (fixed)    &   4884 (fixed)   &   5960 (fixed)      &   5499 (fixed)      &  5150 (fixed)     \\
 $T_2$ [K]                    &  11127 $\pm$ 1109     & 10831 $\pm$ 736   &  15817 $\pm$ 468  &  23425 $\pm$ 312  &   4691 $\pm$ 208 &   5961 $\pm$ 145    &   5125 $\pm$ 98     &  5127 $\pm$ 47    \\
 $R_1/a$                      &  0.277 $\pm$ 0.006    & 0.276 $\pm$ 0.006 & 0.371 $\pm$ 0.005 & 0.290 $\pm$ 0.005 & 0.400 $\pm$ 0.003&  0.389 $\pm$ 0.003  &   0.374 $\pm$ 0.002 & 0.371 $\pm$ 0.002 \\
 $R_2/a$                      &  0.206 $\pm$ 0.010    & 0.241 $\pm$ 0.009 & 0.362 $\pm$ 0.005 & 0.278 $\pm$ 0.008 & 0.360 $\pm$ 0.003&  0.401 $\pm$ 0.003  &   0.370 $\pm$ 0.002 & 0.342 $\pm$ 0.002 \\
 $L_1$ [\%]                   &  31.4 $\pm$ 2.5       &  35.1 $\pm$ 1.9   &  26.2 $\pm$ 1.2   & 30.7 $\pm$ 1.9    &  42.6 $\pm$ 1.0  &  32.1 $\pm$ 1.3     &  30.4 $\pm$ 0.6     & 33.3 $\pm$ 0.7    \\
 $L_2$ [\%]                   &   7.0 $\pm$ 1.7       &  16.4 $\pm$ 2.7   &  17.1 $\pm$ 1.5   & 26.5 $\pm$ 1.3    &  27.4 $\pm$ 0.8  &  28.4 $\pm$ 1.9     &  19.3 $\pm$ 0.5     & 28.3 $\pm$ 0.5    \\
 $L_3$ [\%]                   &  61.6 $\pm$ 5.3       &  48.5 $\pm$ 4.0   &  56.7 $\pm$ 2.3   & 42.8 $\pm$ 2.8    &  30.0 $\pm$ 1.1  &  39.5 $\pm$ 3.0     &  50.3 $\pm$ 0.9     & 38.4 $\pm$ 1.0    \\ \hline
  \multicolumn{1}{c|}{ }      & \multicolumn{7}{c}{{\sc p a i r \,\, B}}\\ \hline
 $i$ [deg]                    &   84.57 $\pm$ 1.23    &  87.77 $\pm$ 1.56 &  46.09 $\pm$ 0.86 & 82.68 $\pm$ 0.80  & 78.63 $\pm$ 0.25 &  84.67 $\pm$ 0.19   &  88.39 $\pm$ 0.33   & 68.82 $\pm$ 1.0    \\
 $q=\frac{M_2}{M_1}$          &  1.00 (fixed)         & 1.00 (fixed)      &   1.00 (fixed)    &  1.00 (fixed)     &  1.00 (fixed)    &   1.00 (fixed)      &  1.00 (fixed)       &  1.04 $\pm$ 0.02   \\
 $T_1$ [K]                    &  15123 (fixed)        & 11558 (fixed)     &  20777 (fixed)    &  24500 (fixed)    &   4884 (fixed)   &   5960 (fixed)      &  5499 (fixed)       &  5150 (fixed)      \\
 $T_2$ [K]                    &  16803 $\pm$  1002    &  7762 $\pm$ 435   &  19818 $\pm$ 359  &  22949 $\pm$ 805  &   4733 $\pm$ 78  &   4380 $\pm$ 119    &  5417 $\pm$ 79      &  5148 $\pm$ 101    \\
 $R_1/a$                      &  0.075 $\pm$ 0.005    & 0.243 $\pm$ 0.009 & 0.370 $\pm$ 0.005 & 0.160 $\pm$ 0.006 & 0.174 $\pm$ 0.002&  0.063 $\pm$ 0.001  &  0.088 $\pm$ 0.001  &  0.392 $\pm$ 0.002 \\
 $R_2/a$                      &  0.090 $\pm$ 0.007    & 0.098 $\pm$ 0.004 & 0.373 $\pm$ 0.007 & 0.157 $\pm$ 0.004 & 0.178 $\pm$ 0.003&  0.063 $\pm$ 0.002  &  0.084 $\pm$ 0.002  &  0.403 $\pm$ 0.002 \\
 $L_1$ [\%]                   &  19.2 $\pm$ 0.8       & 43.4 $\pm$ 1.6    &   27.8 $\pm$ 0.8  & 22.6 $\pm$ 2.0    &  17.2 $\pm$ 0.6  &  34.8 $\pm$ 0.3     &   25.8 $\pm$ 0.4    &  15.0 $\pm$ 0.4    \\
 $L_2$ [\%]                   &  37.0 $\pm$ 1.3       &  4.2 $\pm$ 0.7    &   25.9 $\pm$ 0.9  & 20.3 $\pm$ 1.4    &  16.0 $\pm$ 1.1  &   6.0 $\pm$ 0.6     &   24.4 $\pm$ 0.5    &  15.1 $\pm$ 0.5    \\
 $L_3$ [\%]                   &  43.8 $\pm$ 2.6       & 52.4 $\pm$ 4.2    &   46.2 $\pm$ 2.8  & 57.1 $\pm$ 4.2    &  67.0 $\pm$ 2.0  &  59.2 $\pm$ 0.9     &   49.8 $\pm$ 1.0    &  69.9 $\pm$ 1.6    \\
 \noalign{\smallskip}\hline
\end{tabular}} \\
  %\begin{flushleft}
  %\footnotesize Note: Individual uncertainties of parameters taken from {\sc PHOEBE} only, which are usually underestimated.\\
  %\end{flushleft}
\end{table*}

\begin{table*}%[h!]
\caption{Results of the combined analysis of the ETV data for both A and B pairs.}
 \label{TabETV}
% \scriptsize
 \tiny
  \centering \scalebox{0.645}{
\begin{tabular}{c | c c c c c c c c }
   \hline\hline %\noalign %{\smallskip}
 \multicolumn{1}{c|}{    } &   OGLE SMC-ECL-2339       &   OGLE SMC-ECL-3075       &    OGLE SMC-ECL-4756      &    OGLE SMC-ECL-6093      &    GSC 01949-01700        &   ZTF J171602.61+273606.5  &    WISE J210935.8+390501   & V597 And \\ \hline
  $JD_{0,A}$ [HJD-2450000] & 5000.291 $\pm$ 0.002      & 5000.122 $\pm$ 0.001      &  8326.759 $\pm$ 0.002     & 5001.161 $\pm$ 0.005      & 9538.254 $\pm$ 0.001      &  9440.075 $\pm$ 0.002      &  8966.154 $\pm$ 0.001      & 9862.816 $\pm$ 0.001 \\
 $P_A$ [day]               & 0.7288358 $\pm$ 0.0000005 & 1.3588955 $\pm$ 0.0000007 & 0.9177287 $\pm$ 0.0000004 & 0.9019310 $\pm$ 0.0000020 & 0.2405835 $\pm$ 0.0000001 &  0.3600139 $\pm$ 0.0000003 & 0.3322802 $\pm$ 0.00000001 & 0.4676964 $\pm$ 0.0000003 \\
  $JD_{0,B}$ [HJD-2450000] & 5000.611 $\pm$ 0.002      & 5000.615 $\pm$ 0.009      & 5001.459 $\pm$ 0.002      & 5001.212 $\pm$ 0.016      & 9565.342 $\pm$ 0.001      &  9729.469 $\pm$ 0.007      &  8966.180 $\pm$ 0.001      & 9862.641 $\pm$ 0.001 \\
 $P_B$ [day]               & 3.3957619 $\pm$ 0.0000022 & 2.4158653 $\pm$ 0.0000078 & 2.0604735 $\pm$ 0.0000004 & 2.0303269 $\pm$ 0.0000317 & 0.7583381 $\pm$ 0.0000005 &  4.5154530 $\pm$ 0.0000106 & 3.5157541 $\pm$ 0.0000013  & 0.3525007 $\pm$ 0.0000003 \\
 $p_{AB}$ [yr]             &  5.95 $\pm$ 0.31          &      9.75 $\pm$ 1.84      &   4.34 $\pm$ 0.16         &  31.2 $\pm$ 8.0           &  21.7 $\pm$ 0.5           &   19.5 $\pm$ 0.5           &     1.9 $\pm$ 0.1          & 20.4 $\pm$ 0.4  \\
 $A_A$ [d]                 & 0.006 $\pm$ 0.002         &     0.016 $\pm$ 0.002     &   0.016 $\pm$ 0.002       &  0.044 $\pm$ 0.008        & 0.024 $\pm$ 0.003         &  0.041 $\pm$ 0.001         &   0.006 $\pm$ 0.001        & 0.021 $\pm$ 0.002 \\
 $A_B$ [d]                 & 0.007 $\pm$ 0.002         &     0.019 $\pm$ 0.005     &   0.005 $\pm$ 0.002       &  0.047 $\pm$ 0.015        & 0.029 $\pm$ 0.008         &  0.045 $\pm$ 0.003         &   0.007 $\pm$ 0.001        & 0.025 $\pm$ 0.002 \\
 $e$                       & 0.504 $\pm$ 0.028         &     0.356 $\pm$ 0.021     &   0.299 $\pm$ 0.014       &  0.017 $\pm$ 0.006        &  0.279 $\pm$ 0.011        &  0.456 $\pm$ 0.010         &   0.416 $\pm$ 0.009        & 0.621 $\pm$ 0.027 \\
 $\omega$ [deg]            & 109.2 $\pm$ 29.1          &     344.2 $\pm$ 11.8      &    22.1 $\pm$ 7.9         &  341.4 $\pm$ 17.8         &  160.6 $\pm$ 2.7          &  300.3 $\pm$ 10.6          &   308.9 $\pm$ 6.2          & 24.7 $\pm$ 13.2   \\
 $T_0$ [HJD-2450000]       & 8360.3 $\pm$ 141.4        &    5726.3 $\pm$ 871.0     &   5805.3 $\pm$ 108.2      &  6794.4 $\pm$ 2010.7      &  1789.2 $\pm$ 209.8       &  9836.4 $\pm$ 294.0        &   9861.4 $\pm$ 54.6        & 9436.6 $\pm$ 129.9   \\
 \noalign{\smallskip}\hline
\end{tabular}} \\
\end{table*}

\subsection{ZTF J171602.61+273606.5}

The star ZTF J171602.61+273606.5 was discovered in the ZTF survey as an eclipsing binary by
\cite{2020ApJS..249...18C}. No detailed analysis was published.

Our analysis led to the findings in Figures \ref{FigLCOC} and results given in Tables \ref{TabLC} and
\ref{TabETV}. A slightly asymmetric pair A was found; however, we did not use any spot hypothesis here.
Pair B shows only a very shallow secondary eclipse, indicating a low luminosity of such a component.
The predicted angular separation resulted in about 24 mas here.

\subsection{WISE J210935.8+390501}

The system named WISE J210935.8+390501 was detected as an eclipsing binary thanks to the WISE data (see
\citealt{2018ApJS..237...28C}). Only the shorter orbital period of pair A with a period of about
0.33228 days was discovered, despite the fact that pair B shows even deeper eclipses.

Our analysis as presented below revealed that both A and B pairs are circular, and their masses are also similar to each other (both regarding the similar luminosities of both pairs and the similar amplitudes in the ETV diagrams). Here, we deal with the shortest mutual period of the quadruple, which is only approximately two years long. Thanks to this fact, we have several outer periods covered with data today, and the bound quadruple 2+2 hypothesis is definitely proven.   %  predicted angular.separ 7mas

\subsection{V597 And}

The last system in our sample is named V597 And. It is also probably the most interesting one. This
star was incorrectly listed as an RR Lyr star in Simbad according to \cite{2007PZ.....27....2D}, who
gave a period of about 0.2338 days. Its correct period is double this, and it is definitely an
eclipsing type. This is the brightest star in our sample.

We followed more or less the same procedure as for the previous systems, using the TESS data as the best for the LC fitting. There appears that the LC shape of pair A shows asymmetry, probably due to surface spots. However, the different sectors of TESS data show a slightly different shape of the LC; hence, some spot evolution can be traced here. % This system shows the shortest orbital periods of A \& B pairs.
This complication makes the analysis more difficult; however, its long-term evolution of both the
orbital periods is easily detectable here. This is especially true thanks to available data from the
older photometric databases such as ASAS-SN, SuperWASP, Atlas, and KWS. The star has deep enough
eclipses and a bright enough magnitude that both pairs were detectable in the older data. Despite the
mutual A-B period of about 20 years, it is well covered with regards to data today, and its 2+2
quadruple nature has been proven.

There appears to be one problem with the most distant data point in our Figure \ref{FigLCOC} near the
year  2000, which comes from the published observation of the star in the GCVS catalogue
\citep{2017ARep...61...80S}. We are not able to identify this one observation with either pair A or
pair B. Therefore, we mark this data point with a question mark in Figure \ref{FigLCOC} and did not
take it into account in our analysis. Assuming that it was the observation of pair A, it should
indicate a longer mutual orbital period. On the contrary, if it belongs to pair B, then the ETV curve
needs to be shorter or have a higher amplitude. We leave this as an open question since the star was
classified as RR Lyr originally, so there is still a possibility that this observation indicates a
maximum and not a minimum brightness.

The system is also closest to the Sun, according to GAIA DR3 \citep{2023A&A...674A...1G} only about 400
pc distant. Thanks to this value, the predicted angular separation of the double should be of about 57
mas. Such a separation is within the capabilities of the current technique, and the high-resolution
methods should be used to resolve both components. Finally, the star was also found to contain probably
another close faint component at about 3.6$^{\prime\prime}$. In the GAIA DR3 catalogue both the
components share the same parallax and proper motion. The result is that this system is a rare
quintuple system of (2+2)+1 architecture.

 \medskip

\section{Conclusions}  \label{discussion}

We performed the very first analysis of eight new quadruple systems and proved their gravitational
coupling. All of them are now proven to constitute rare 2+2 quadruples thanks to the long-term analysis
of their ETV curves for both pairs. This was especially possible due to older photometric data from
various databases, as well as thanks to our new dedicated observations of these systems.

All of the systems presented here are new detections; that is, they were not listed as candidate doubly
eclipsing systems before. Most of the stars have their mutual A-B orbital periods of the order of years
to decades. Such periods are nowadays short enough to be well covered with reliable observations. On
the other hand, they are long enough that the dynamical interactions of the two doubles are small
enough to be almost negligible. For the most compact system (i.e. this one, where $P_{AB}$/$P_A$, or
$P_{AB}$/$P_B$ is the smallest),  WISE J210935.8+390501, the period of these long-term perturbations
($P_{AB}^2$/$P_A$, or $P_{AB}^2$/$P_B$, see e.g. \citealt{2015MNRAS.448..946B}), is several centuries
long. Hence, it can definitely be ignored in our dataset spanning a few decades at maximum.

However, one can ask whether it is still important to discover such systems of 2+2 architecture and
derive their physical and orbital properties. Ten years ago, this was new, and only a few such 2+2
stars were known at that time. Nowadays, we have several dozen of them, with new ones being discovered
every year. Besides filling the statistics these systems can help us to understand their formation
mechanisms, indicating that there are probably different formation routes for the 2+1+1 and for 2+2
quadruples \citep{2021Univ....7..352T}. Moreover, such complicated tight multiples can sometimes offer
us a possibility of detecting effects that are otherwise undetectable \citep{2023Univ....9..485B}.

\begin{acknowledgements}
%An anonymous referee is being acknowledged for useful comments and suggestions, greatly improving the manuscript.
 We do thank the SuperWASP, ZTF, ASAS-SN, NSVS, KWS, CRTS, Atlas, OGLE, MACHO, and TESS teams for making all of the observations easily public available.
 We are also grateful to the ESO team at the La Silla Observatory for their help in maintaining and operating the Danish 1.54m telescope.
 I.\v{S}\'andorov\'a is acknowledged for some preliminary reduction of the data.
  This work has made use of data from the European Space Agency (ESA) mission {\it Gaia} (\url{https://www.cosmos.esa.int/gaia}), processed by the {\it Gaia}
Data Processing and Analysis Consortium (DPAC,
\url{https://www.cosmos.esa.int/web/gaia/dpac/consortium}). Funding for the DPAC has been provided by
national institutions, in particular the institutions participating in the {\it Gaia} Multilateral
Agreement.
 The research of P.Z., J.K., and J.M. was also supported by the project {\sc Cooperatio - Physics} of Charles University in Prague.
 The observations by Z.H. in Velt\v{e}\v{z}e were obtained with a CCD camera kindly borrowed by the Variable Star and Exoplanet Section of the Czech Astronomical Society.
 This research made use of Lightkurve, a Python package for TESS data analysis \citep{2018ascl.soft12013L}.
 This research has made use of the SIMBAD and VIZIER databases, operated at CDS, Strasbourg, France and of NASA Astrophysics Data System Bibliographic Services.
\end{acknowledgements}

\end{document}